\documentclass{PoS}

\title{A High Resolution Neutrino Experiment in a Magnetic Field for Project-X at Fermilab}

\ShortTitle{New Experiments: HiResM$\nu$}

\author{SANJIB R. MISHRA \\
        Department of Physics and Astronomy, University of South Carolina, Columbia SC 29208, USA \\  
        E-mail: \email{sanjib@sc.edu}}

\author{\speaker{ROBERTO PETTI}%
        \\
        Department of Physics and Astronomy, University of South Carolina, Columbia SC 29208, USA \\  
        E-mail: \email{roberto.petti@cern.ch}}

\author{CARL ROSENFELD\\
        Department of Physics and Astronomy, University of South Carolina, Columbia SC 29208, USA \\  
        E-mail: \email{lcr@sc.edu}}

\abstract{   
We propose a new high-resolution neutrino experiment within a
dipole magnetic field, HiResM$\nu$. This experiment will run along
with long-baseline neutrino oscillation experiments (LBL$\nu$)
such as NO$\nu$A, a large-cavity detector at DUSEL, or a Liquid-Argon detector
in the Medium-Energy (ME) configuration of the NuMI-beam.
The $4 \times 4 \times 7$~m$^3$ detector, inside a
dipole magnetic field of \boldmath{$B \approx 0.4$~T},
will have the density of liquid hydrogen,
$\rho \approx 0.1$~gm/cm$^3$, with a nominal fiducial mass of
7.4~tons. Assuming the 120~GeV Main Injector proton intensities 
we anticipate 140(50) million $\nu_\mu$ ($\overline{\nu}_\mu$) 
Charged-Current (CC) events in the fiducial volume, for 3(4)-year 
run with the ME (anti)neutrino beam. Alternatively, the same statistics 
could be collected in just 1(1.5) year with the High Energy (HE) beam configuration.  
The goals of HiResM$\nu$ are twofold. It will measure the relative abundance 
and the energy spectrum of $\nu_\mu$, $\overline{\nu}_\mu$, $\nu_e$ and $\overline{\nu}_e$
CC-interactions, which are directly relevant to LBL$\nu$.    
It will serve as an `Event-Generator'  of real neutrino interactions to estimate backgrounds and efficiencies in LBL$\nu$. As such, it  will provide a quantitative determination of the overall energy-scale of neutrino CC interactions and of hadronic multiplicities for all CC and Neutral-Current (NC) event topologies.  
In addition, it will perform precise measurements of electroweak parameters, 
of (semi)exclusive processes such as quasi-elastic, resonance, $\pi^0$/$K^0_S$/$\Lambda$/charm-hadron
production, as well as of the hadronic structure of nucleons and nuclei.   
We expect to reach a sensitivity of about 0.2\% on the weak mixing angle 
by combining different channels. The new experiment will also perform 
searches for new physics beyond the Standard Model.   
}

\FullConference{10th International Workshop on Neutrino Factories, Super beams and Beta beams\\
		 June 30 - July 5 2008\\
		 Valencia, Spain}

\begin{document}

\section{High Intensity Neutrino Beam}

A new high intensity proton facility, Project-X, is planned to support a major program in neutrino 
and flavor physics at Fermilab~\cite{projectX}. Project X is based on a new 8 GeV superconducting   
linac, paired with the existing (but modified) Main Injector and Recycler Ring, to provide in excess 
of 2 MW of beam power throughout the energy range 60-120 GeV. The linac utilizes technology in common 
with the International Linear Collider over the energy range 0.6-8.0 GeV. The project would allow to 
achieve an intensity of $1\times 10^{18}$~protons/hour for the Main Injector 120~GeV beam.   
Assuming (365/$\pi$) days of operation in a year, that corresponds to about 30$\times 10^{20}$ protons/year available for the MI neutrino beam.   
  
We propose a new neutrino experiment, HiResM$\nu$, which, taking advantage of the unprecedented (anti)neutrino fluxes available with Project-X, should combine high resolution 
in the reconstruction of neutrino events and large statistics.  
The experiment will run along with the LBL$\nu$ experiments at the Near Detector site. 
This fact is an imperative to achieve the
highest precision in the discovery of the elements of the neutrino mass matrix.
The LBL$\nu$'s will run in both neutrino and anti-neutrino modes. Assuming the Medium 
Energy spectrum of the existing NuMI beam and a modest fiducial mass of 7.4 tons, 
we plan to collect 140 million $\nu_\mu$ CC events and 50 million $\overline{\nu}_\mu$ CC 
events in a 3-year and 4-year run, respectively. It is worth noting the running time required 
to collect the same statistics would be reduced by about a factor of three with the 
High Energy configuration of the NuMI beam.

\section{High Resolution Detector}

\begin{figure}[p!]
\begin{center}
\hspace*{-0.50cm}\includegraphics[scale=0.49,angle=0]{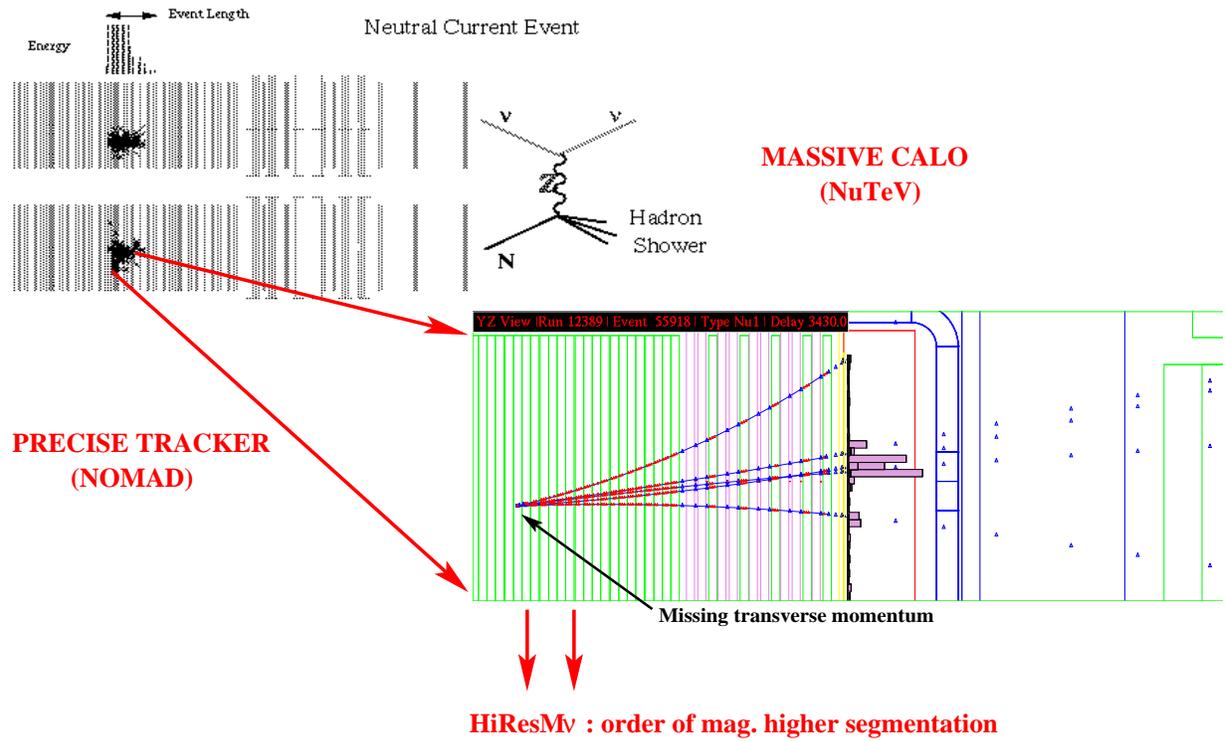}
\caption{Candidate NC Event in NuTeV and NOMAD.
In tracking charged particles
HiResM$\nu$ will provide a factor of two higher segmentation
along $z$-axis and a factor of six higher segmentation
in the transverse-plane compared to NOMAD.}
\label{fig-nc-nutev-nomad}
\end{center}
\end{figure}

Building upon the NOMAD-experience~\cite{NOMAD},
we propose a low-density tracking detector as neutrino target. 
The active target tracker will have a factor of two more sampling 
points along the $z$-axis ($\nu$-direction) and a factor of six more sampling points
in the plane transverse to the neutrino compared to the NOMAD experiment.
Figure~\ref{fig-nc-nutev-nomad} juxtaposes the resolving power of the NOMAD detector with the massive
CCFR/NuTeV calorimeter. One sees a stark contrast for an NC event candidate in the NuTeV experiment
compared with one in the NOMAD experiment. The proposed experiment, HiResM$\nu$, will further 
enhance the resolving power: an order of magnitude more points in tracking charged particles, and
coverage for side-exiting neutrals and muons.

Taking advantage of the existing design and production details for the
ATLAS Transition Radiation Tracker~\cite{ATLAS-TRT}
and the COMPASS detector~\cite{COMPASS-straw}, we are proposing straw-tube 
trackers (STT) for the active neutrino target of HiResM$\nu$. 
The tracker will be composed of straw tubes with 1~cm diameter. Vertical ($Y$) and horizontal ($X$) straws
will be alternated and arranged in modules. In front of each module a plastic radiator made of many thin foils
allows the identification of electrons through their Transition Radiation. 
The nominal fiducial volume (FV) for CC analysis is: 
$350 \times 350 \times 600$~cm$^3$, corresponding to 7.4 tons of mass 
with an overall density $\rho \leq 0.1$~gm/cm$^3$. The STT  will be surrounded by 
an electromagnetic calorimeter (sampling Pb/scintillator) covering the forward and side regions.   
Both sub-detectors will be installed inside a dipole magnet providing a magnetic field 
of $\sim 0.4$ T. An external muon detector based upon Resistive Plate Chambers (RPC) will 
be placed outside of the magnet.   

The neutrino target would be mainly composed of carbon, and a radiation length is about 5 $m$.
The spacepoint resolution is expected to be $<200 \mu m$, consistent with the performance of 
ATLAS~\cite{ATLAS-TRT} and 
COMPASS~\cite{COMPASS-straw} trackers. Multiple scattering contributes 0.05 to the 
$\Delta p/p$ for tracks 1 $m$ long, while the measurement error would be 0.006 for $p=1$ GeV tracks.
The proposed detector will  measure track position, $dE/dx$, and Transition Radiation (with Xe filling)
over the entire instrumented volume. The unconverted photon energy will be measured in the calorimeters 
with a target energy resolution of $\sim 10\% / \sqrt{E}$. The expected capabilities of
HiResM$\nu$ include:

\begin{itemize}
\vspace{-0.20cm}
\item Full reconstruction of charged particles and $\gamma$'s;
\vspace{-0.20cm}
\item Identification of $e$, $\pi$, $K$, and $p$ from $dE/dx$;
\vspace{-0.20cm}
\item Electron (positron) identification from Transition Radiation ($\gamma > 1000$);
\vspace{-0.20cm}
\item Full reconstruction and identification of protons down to momenta of 250 MeV;
\vspace{-0.20cm}
\item Reconstruction of electrons down to momenta of 80 MeV from curvature in $B$ field.
\vspace{-0.20cm}
\end{itemize}

The proposed design provides both redundancy of measurements and flexibility. The 
redundancy is crucial for high resolution in the reconstruction of neutrino events.
Furthermore, most of the target mass (85\%) is represented by the radiators,
which are independent from the straws. This fact allows a change in the fiducial  
mass without affecting the construction of the trackers.

\thispagestyle{empty}

\begin{figure}[p!] 
\begin{center}
\includegraphics[scale=0.51]{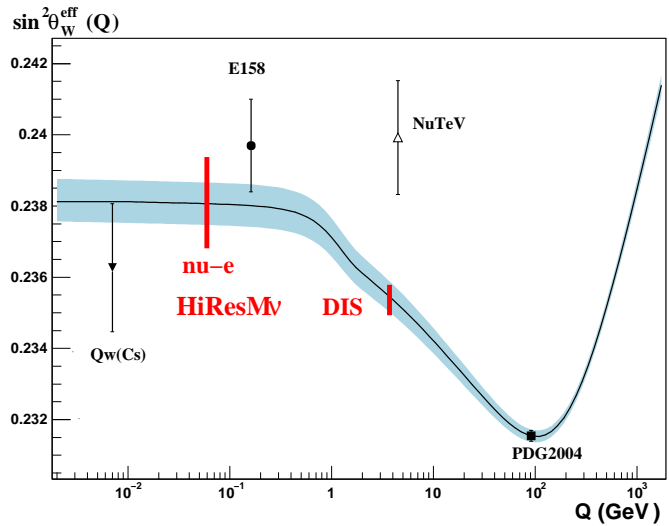}
\caption{Running of the weak mixing angle as a function of
the momentum transfer $Q$, as predicted by the Standard Model~\cite{marciano00}.
The data points are from Atomic Parity Violation~\cite{APV-sin2w}, Moeller
scattering (E158~\cite{E158-sin2w}), $\nu$ DIS (NuTeV~\cite{nutev-sin2w})
and the combined $Z$ pole  measurements (LEP/SLC).
The projected sensitivity of our experiment is shown for comparison. }
\label{fig-wma-run}
\end{center}
\end{figure}

\section{Reduction of Systematics for LBL$\nu$}

The sensitivity to the oscillation parameters ($\theta_{13},\delta_{\rm CP}$) 
which can be achieved by the next generation LBL$\nu$ experiments will  
depend upon systematic uncertainties from intrinsic flux contaminations   
and from the knowledge of (anti)neutrino cross-sections and background 
topologies. The proposed experiment will measure the relative abundance,
the energy spectrum, and the detailed topologies
for $\nu_\mu$/$\overline \nu_\mu$/$\nu_e$/$\overline \nu_e$
induced interactions including the momentum vectors of negative, positive, and
neutral ($\pi^0$ and $K^0_s$/$\Lambda$/$\overline \Lambda$)
particles composing the hadronic system.
The experiment will provide topologies, on an event-by-event basis, of various interactions
that will serve as `generators' for the simulation of LBL$\nu$ experiments.
A glance at the NC event candidate in NOMAD, shown in Figure~\ref{fig-nc-nutev-nomad}, 
gives an idea of the precision with which the charged-particles and the forward-$\gamma$ 
were measured. The proposed experiment will have substantially better resolution than NOMAD.

The excellent resolution of the detector inside a
calibrated dipole {\bf B}-field will allow a precise
determination of the overall neutrino energy scale, a crucial scale in
precisely determining the neutrino oscillation parameters in
LBL$\nu$. The energy scale of charged particles and the B-field calibration
can be checked from the mass constraint of the reconstructed $K^0_S$.
With a total sample of 30,000 reconstructed $K^0_S$ in NOMAD the charged track energy scale 
was determined to better than 0.2\%. Similarly, the hadron energy scale
was constrained to  0.5\% level from the reconstructed charged
tracks and from the muon measurement. Both uncertainties were limited
by the statistics and resolution of the available control samples~\cite{NOMAD-xsec}.
In the proposed experiment, we expect to reconstruct over $2 \times 10^{6}$ $K^0_S$
in the $\nu$-mode  and we will collect overall a factor of 200 more 
($\nu + \bar{\nu}$) events than in NOMAD.

Because this experiment and the LBL$\nu$ experiments will
utilize the same neutrino beam, our irreducible errors in
measured cross sections, in the NC/CC ratio, in the species
composition of hadronic secondaries, and in the
($x_F, P_T$) of hadronic secondaries will not propagate to
the determination  of the oscillation parameters in the LBL$\nu$.

\section{Electroweak Measurements}

One goal of the proposed  experiment is to
measure the weak mixing angle, sin$^2\theta_W$,
with a precision approaching 0.2\%, i.e.
$\delta$sin$^2\theta_W \approx \pm 0.00045$ (on-shell),
a precision comparable to the PDG value~\cite{PDG08}.
The  current PDG precision on sin$^2\theta_W$ derives from the
LEP/SLC/CDF/D0 measurements. The proposed
$\nu$-experiment, the only direct probe to
$\nu$-Z coupling, aims to  measure this quantity at values of
$Q^2$ that are 1/1000 of those at colliders with
commensurate precision. Finally, the NuTeV experiment
has reported an anomalous value of sin$^2\theta_W$ that is
about 3$\sigma$ higher than the `world average'~\cite{nutev-sin2w}.
The HiResM$\nu$ will provide a decisive check of this anomaly.

In HiResM$\nu$, two different channels permit precise measurements of
sin$^2\theta_W$ with independent systematics and at at different
scales ($Q^2$), as shown in Figure~\ref{fig-wma-run}.
The most promising channel for the sin$^2\theta_W$ measurement
is the Deep Inelastic Scattering (DIS) with a precision approaching 0.2\%. It is followed
by the $\nu$-e scattering where a  precision of 0.56\% can be achieved.

\subsection{Deep Inelastic Scattering off quarks} 

The experiment will permit a measurement of
R$^\nu$, the NC/CC ratio in a $\nu$-beam, and of
R$^{\overline \nu}$, the NC/CC ratio in a $\overline \nu$-beam.
We will also exploit the Paschos-Wolfenstein relation~\cite{pw-R-} to
constrain the systematic errors.
With a cut on the hadronic energy E$_{Had} \geq 3$~GeV, the
NC and the CC samples in the NuMI-ME beam are almost entirely composed of
deep-inelastic scattering (DIS) events. The number of NC events
in the $\nu$-mode is 19$\times 10^6$, and that in the
$\overline \nu$-mode is 4$\times 10^6$. Thus, the expected statistical 
uncertainty on sin$^2\theta_W$ is $0.08\%$.   

The total experimental errors on R$^\nu$ (R$^{\overline \nu}$)
in the proposed experiment will be a factor of 4 smaller
than those quoted for the NuTeV-experiment~\cite{nutev-sin2w}
due to the higher resolution and statistics. The overall expected 
experimental systematics on sin$^2\theta_W$ is $0.1\%$, mainly 
dominated by the muon identification and by the kinematic analysis 
like in the NOMAD experiment. For $R^\nu$ measurement, flux is almost a non-issue.
For the $R^-$, we will constrain the energy-integrated
$\overline \nu_\mu / \nu_\mu$ flux ratio using the Low-$\nu^0$ method of determining
relative flux~\cite{Mishra-nu0} in concert with the measurements by MIPP-II and NA49.
We will also use the knowledge of the extant $\overline \nu_\mu$-CC and $\nu_\mu$-CC  data.
Such an effort will yield a precision of the order of 1\% on the cumulative
$\overline \nu_\mu / \nu_\mu$ flux.

The theoretical uncertainty in the NuTeV measurement is dominated by
errors from charm production and strange sea followed by the errors
from the longitudinal structure function (F$_L$) and higher-twist effects.
First, there are $\sim$ 15,000 charm-induced dimuons
in the NOMAD experiment (on-going analysis). A global analysis of the charm 
production and strange sea, including the NOMAD, CCFR, and NuTeV data, will reduce the
NuTeV error by a factor of 2. Importantly, we expect to measure $in$ $situ$
$\sim$ 200,000 charm-induced dileptons ($e^+$ and $\mu^+$). 
The detector will also permit direct measurements of exclusive charmed hardons 
($\sim$ 5 million), which will be reconstructed  from their decay kinematics.  
The estimate of the other theoretical uncertainties is based upon the
current understanding of neutrino-nucleus structure functions
~\cite{AKP,EWRC,KP}, and the on-going sin$^2\theta_W$ analysis in NOMAD ($E_{Had} \geq 3$ GeV).
The theoretical models used for our calculations are anchored in the fits to the existing experimental 
data. These models deal with higher twist, longitudinal structure function,
bound state effects in nuclei and electroweak corrections.
Some of the uncertainties will be further reduced by J-Lab measurements,
and measurements conducted in HiResM$\nu$ experiment.
We expect the model uncertainty on sin$^2\theta_W$ to be about $0.14\%$.

\subsection{Elastic Scattering off electrons} 

The $\nu$-$e^-$ scattering, via CC or NC, results in a clean
signal in this experiment. The signal is a single $e^-$ ($\mu^-$)
emerging at zero-angle in the NC (CC) reaction.
The relevant measureable that characterizes the signal
is $E_L \theta_L^2$ where
$E_L$ and $\theta_L$ are the energy and angle (with respect
to the neutrino direction) of the emergent lepton.
Thus, the key to measuring the $\nu$-$e^-$ NC interaction is the
presence of B-field and resolution in $\theta_e$.
The background, almost entirely caused by
photon conversion, is charge independent. Thus, we will
measure the background to the $\nu$-$e^-$ NC process by
measuring forward $e^+$. Moreover, the energy and the angle of the scattered 
$e^-$ will be measured with high resolution in HiResM$\nu$.

We posit that the dominant error will be  the statistical error of
the $\nu$-$e^-$ sample. The ratio of $\sigma (\overline \nu_\mu e^-)$ and $\sigma (\nu_\mu e^-)$
gives a measure of sin$^2\theta_W$.  
Since we aim to measure $R_{\nu e^-}$ to a relative precision of about
1.0\%, the corresponding error on sin$^2\theta_W$ will be about 0.56\%.


\begin{thebibliography}{99}

\bibitem{projectX} 
        \begin{verbatim}http://www.fnal.gov/pub/directorate/steering/pdfs/SGR_2007.pdf \end{verbatim}  
        \vspace*{-0.45cm}  
        \begin{verbatim}http://www.fnal.gov/directorate/Longrange/Steering_Public/P5.html \end{verbatim} 

\bibitem{NOMAD} J. Altegoer et al. [NOMAD collaboration],  
                Nuclear Instruments and Methods A 404 (1998) 96-128;
                P. Astier et al. [NOMAD collaboration], 
                Nucl. Phys. B611 (2001) 3-39, hep-ex/0106102;   
                P. Astier et al. [NOMAD collaboration], 
                Phys. Lett. B 570 (2003) 19-31, hep-ex/0306037 

\bibitem{NOMAD-xsec} Q. Wu et al. [NOMAD collaboration],
                     Phys. Lett. B 660 (2008) 19-25; arXiv:0711.1183 [hep-ex]

\bibitem{ATLAS-TRT} T. Akesson et al., Nucl.Instrum.Meth. A522 (2004) 131-145;  
                    Nucl.Instrum.Meth. A522 (2004) 50-55;  
                    IEEE Nucl.Sci.Symp.Conf.Rec.2 (2005) 1185-1190; 
                    E. Abat et al., JINST 3 (2008) P02013; JINST 3 (2008) P02014.   

\bibitem{COMPASS-straw} V.N. Bychkov et al., Particles and Nuclei Letters, 2 (111) June 2002;  
                        K. Platzer et al., IEEE Transactions on Nuclear Science, vol. 52, No. 3, June 2005.   

\bibitem{nutev-sin2w} G.~P.~Zeller et al. [NuTeV collaboration], Phys. Rev. Lett. {\bf 88}, 091802 (2002), arXiv:hep-ex/0110059.  


\bibitem{pw-R-}  E.~A.~Paschos and L.~Wolfenstein,  
                 Phys.\ Rev.\  D {\bf 7}, 91 (1973).

\bibitem{marciano00} A. Czarnecki and W.J. Marciano, 
                     Int. J. Mod. Phy. A15 (2000) 2365  

\bibitem{APV-sin2w} S.C. Bennet and C.E. Wieman, Phys. Rev. Lett. 82 (1999) 2484 

\bibitem{PDG08} C.~Amsler et al. [Particle Data Group],
                Phys.\ Lett.\  B {\bf 667} (2008) 1.

\bibitem{E158-sin2w} P.L. Anthony et al [E158 collaboration], 
                     Phys.Rev.Lett. 95 (2005) 081601, hep-ex/0504049

\bibitem{AKP} S. Alekhin, S.A. Kulagin and R. Petti, 
              arXiv:0812.4448 [hep-ph]; S. Alekhin, S.A. Kulagin and R. Petti,
              AIP Conf.Proc. 967 (2007) 215-224, arXiv:0710.0124 [hep-ph];   
              S.~Alekhin, S.~A.~Kulagin and R.~Petti,
              {\it Prepared for 16th International Workshop on Deep-Inelastic Scattering
              and Related Subjects (DIS2008)}, London, UK, 7-11 April 2008, arXiv:0810.4893 [hep-ph].

\bibitem{EWRC} A.~B.~Arbuzov, D.~Y.~Bardin and L.~V.~Kalinovskaya,
               JHEP {\bf 78}, 506 (2005); arXiv:hep-ph/0407203;
               K.~P.~Diener, S.~Dittmaier and W.~Hollik,
               Phys.\ Rev.\ D {\bf 69}, 073005 (2004).

\bibitem{KP} S.A. Kulagin and R. Petti, 
             Nucl. Phys. A 765 (2006) 126-187, hep-ph/0412425;     
             S.A. Kulagin and R. Petti, 
             Phys.Rev. D 76 (2007) 094023, hep-ph/0703033.  

\bibitem{Mishra-nu0} S.R.Mishra,
                {\it Workshop on Hadron Structure Functions and Parton Distributions}, 
                Fermilab, Apr.1990; World Scientific,
                84-123(1990), Ed. D.Geesaman $et$ $al.$;
                Nevis Preprint \# 1426, Jun(1990);



\end{thebibliography}
\end{document}